\documentclass[a4paper,11pt]{article}
\usepackage{pos}
\usepackage{mathtools} 
\usepackage{xcolor}
\usepackage{graphicx}
\usepackage{hyperref}
\usepackage{microtype}
\usepackage{cleveref}
\usepackage{booktabs}
\usepackage{xcolor}
\usepackage{tabu}
\usepackage{siunitx}

\newcommand{\cpcite}[0]{Nir:1999mg,Buras:2001pn}
\newcommand{\transitions}[0]{%
Kim:2005gf,
Christ:2005gi,
Hansen:2012tf,
Briceno:2012yi,
Bernard:2012bi,
Briceno:2014uqa}

\title{Exploring distillation at the $SU(3)$ flavour symmetric point}

\author[a]{Felix Erben}
\author[a]{Maxwell T. Hansen}
\author*[a]{Fabian Joswig}
\author[a]{Nelson Pitanga Lachini}
\author[a]{Antonin Portelli}

\affiliation[a]{Higgs Centre for Theoretical Physics, School of Physics and Astronomy,\\
The University of Edinburgh, Edinburgh EH9 3FD, UK}

\emailAdd{fabian.joswig@ed.ac.uk}

\abstract{In these proceedings we present an exact distillation setup with stabilised Wilson fermions at the $SU(3)$ flavour symmetric point utilising the flexibility of the Grid and Hadrons software libraries. This work is a stepping stone towards a non-perturbative investigation of hadronic $D$-decays, for which one needs to control the multi-hadron final states. As a first step we study two-to-two $s$-wave scattering of pseudoscalar mesons. In particular we examine the reliability of the extraction of finite-volume energies as a function of the number of eigenvectors of the gauge-covariant Laplacian entering our distillation setup.}

\FullConference{%
39th International Symposium on Lattice Field Theory - Lattice2022 \\
8-13 August 2022 \\
Bonn, Germany
}

\begin{document}
\maketitle

\section{Introduction}
The violation of charge conjugation symmetry ($\text{C}$) as well as the violation of this combined with parity ($\text{CP}$) are necessary conditions to explain the matter-antimatter asymmetry in the universe \cite{Sakharov:1967dj}. There are various known sources of $\text{CP}$-violation in the standard model of particle physics\footnote{See e.g.~refs.~\cite{\cpcite} for pedagogical overviews.} but their combined effect appears to be too small to account for the observed asymmetry \cite{Rubakov:1996vz}. Recently the LHCb experiment observed nonzero $\text{CP}$ asymmetry in the decay of charmed hadrons for the first time \cite{LHCb:2019hro} and estimated the difference of the time-integrated asymmetries in $D^0\rightarrow K^-K^+$ and $D^0\rightarrow \pi^-\pi^+$ decays to be
\begin{align}
  \Delta A_\mathrm{CP}=A_\mathrm{CP}(K^-K^+)-A_\mathrm{CP}(\pi^-\pi^+)=(-15.4\pm2.9)\times 10^{-4}\,.
\end{align}
These kinds of decays can be a test case for beyond-the-standard-model dynamics in the up-quark sector but the corresponding theoretical standard model predictions are difficult to compute reliably. (See for example ref.~\cite{Khodjamirian:2017zdu}.)

In these proceedings, we describe our progress towards a first calculation of hadronic $D$-decays from first principles using Monte Carlo simulations of lattice QCD at heavier-than-physical quark masses. Our strategy to obtain the desired decay amplitude is to compute the required matrix element via an effective four-quark Hamiltonian $\mathcal{H}_\mathrm{weak}$ \cite{Buchalla:1995vs} in a finite spacetime volume. We can then relate the finite-volume matrix element to its infinite-volume counterpart via
the relation due to Lellouch and L{\"u}scher \cite{Lellouch:2000pv} and subsequent generalizations \cite{\transitions}:
\begin{align}
\label{eq:master_formula}
     |A|^2=8\pi{\bigg\lbrace q\frac{\partial \phi}{\partial q}+k\frac{\partial \delta_0}{\partial k} \bigg\rbrace_{k=k_n}} \frac{E_n^2m_D}{k_{n}^3}\,\big|{Z^{\overline{\mathrm{MS}}}}{\langle} { n,L} {| \mathcal{H}_\mathrm{weak}|D,L\rangle}\big|^2\,,
\end{align}
where the quantity in angled brackets is a finite-volume matrix element that can be determined via lattice QCD, and depends on the box length $L$ and the final state $n$ as indicated. The renormalization factor $Z^{\overline{\mathrm{MS}}}$ is the link between the lattice regularized matrix element and its continuum counterpart. The factor multiplying the renormalized matrix element, often called the Lellouch-L{\"u}scher factor, relates it to the infinite-volume decay amplitude and depends on $\phi(q)$ (a known geometric function of $q = k L /(2 \pi)$ where $k$ is the momentum of the decay products). The factor additionally depends on $\delta_0$, the $s$-wave scattering phase of the final-state particles. A central challenge here is that the $\langle n, L \vert$ state must satisfy $E_n = m_D$, where $m_D$ is the incoming meson mass, in order to define a physical decay.

On the way towards a first full computation of hadronic $D$-decays on the lattice, various theoretical and computational challenges have to be overcome. (See ref.~\cite{Boyle:2022ncb} for a recent more general discussion of numerical challenges in lattice QCD simulations.)
In these proceedings we particularly focus on studying $K \pi$ scattering, with an eye on $D \to K \pi$ decays. The scattering study is required for two reasons. First, as can be seen in eq.~\eqref{eq:master_formula}, the scattering phase shift is required to extract the physical observable. Second, as we detail below, the analysis requires the construction of optimized operators that can then also be used to create the excited $\langle n, L \vert$ state in the decay.

\section{Computational setup}
In this study we work on a set of gauge-field ensembles with three flavors of stabilised Wilson fermions \cite{Francis:2019muy} and tree-level Symanzik improved gluons. The ensembles were generated by the OPEN LATtice initiative \cite{openlattice,Francis:2022hyr,Cuteri:2022gmg} using the openQCD software package \cite{openqcd}.
The three degenerate sea quarks in the simulation are tuned such that the sum of their masses is equivalent to their sum in the physical world. The gauge-field ensembles which we plan to use in this project are summarized in \Cref{tab:gauge_field}. All preliminary results presented in these proceedings are only based on the coarsest ensemble, labeled a12m400. We plan to extend this calculation to two additional ensembles with very similar pion masses and physical volumes but finer lattice spacings, which can be considered to lie on an approximate line of constant physics. This will allow us to control the continuum limit of the calculation which is especially important for observables with heavy valence quarks.
        \begin{table}[ht]
            \centering
            \begin{tabu}{cclclcc}
                \toprule
                Label & $T\times L^3/a^4$ & $\beta$ & $\kappa$ & a (fm) & $m_\pi$ (MeV)\\
                \midrule
                a12m400 & $96\times 24^3$ & 3.685 & 0.1394305 & 0.12 & 410\\
                \midrule
                \rowfont{\color{gray!90!white}}
                a094m400 & $96\times 32^3$ & 3.8 & 0.1389630 & 0.094 & 410\\
                \midrule
                \rowfont{\color{gray!90!white}}
                a064m400 & $96\times 48^3$ & 4.0 & 0.1382720 & 0.064 & 410\\
                \bottomrule
            \end{tabu}
            \caption{Planned gauge-field ensembles for this project. All preliminary results were generated on ensemble a12m400. We plan to include the greyed out ensembles in the near future.}
            \label{tab:gauge_field}
        \end{table}

For the computation of the relevant correlation functions, we make use of the exact distillation method described in ref.~\cite{HadronSpectrum:2009krc}.
In this approach, a smearing matrix $\mathcal{S}$ is obtained from the low-mode subspace of the three-dimensional gauge-covariant Laplacian
\begin{align}
    \mathcal{S}(t)=\sum_{k=1}^{N_\mathrm{vec}}u_k(t)u_k(t)^\dagger\,, \label{eq:smearing_matrix}
\end{align}
where $u_k$ are the eigenvectors of $-\nabla^2$.
Correlation functions can then be cost effectively built from the smeared quark fields
\begin{align}
    \tilde{q}=\mathcal{S}q \,.
\end{align}
For the $I=3/2$ $s$-wave channel, which will be the focus of these proceedings, we construct correlation functions from $K\pi$ two-hadron interpolators with different momenta.

From the relevant operators for a given channel we construct a correlator matrix $C$ and solve a generalized eigenvalue problem (GEVP) \cite{wilson:gevp,Michael:1982gb,Luscher:1990ck,Blossier:2009kd}
\begin{align}
    C(t)v_i(t, t_0)=\lambda_i(t,t_0) C(t_0)v_i(t, t_0)\,, \label{eq:GEVP}
\end{align}
in order to obtain the eigenvalues $\lambda_i(t,t_0) \sim e^{-aE_i(L)(t-t_0)}$ for a desired state $i$.

Our computational setup is based on the \texttt{Grid} \cite{Boyle:2015tjk} and \texttt{Hadrons} \cite{antonin_portelli_2022_6382460} program libraries. The distillation modules are also used in an ongoing $K\pi$ scattering study at the physical point using Domain Wall fermions \cite{Lachini:2021bbr,Lachini:2022xbt}. For the error analysis we make use of the $\Gamma$-method approach \cite{Wolff:2003sm,Ramos:2018vgu} in the \texttt{pyerrors} implementation \cite{Joswig:2022qfe}.

\section{Eigenvector dependence of the finite-volume energies}

\begin{figure}[ht]
        \centering
        \includegraphics[width=0.75\linewidth]{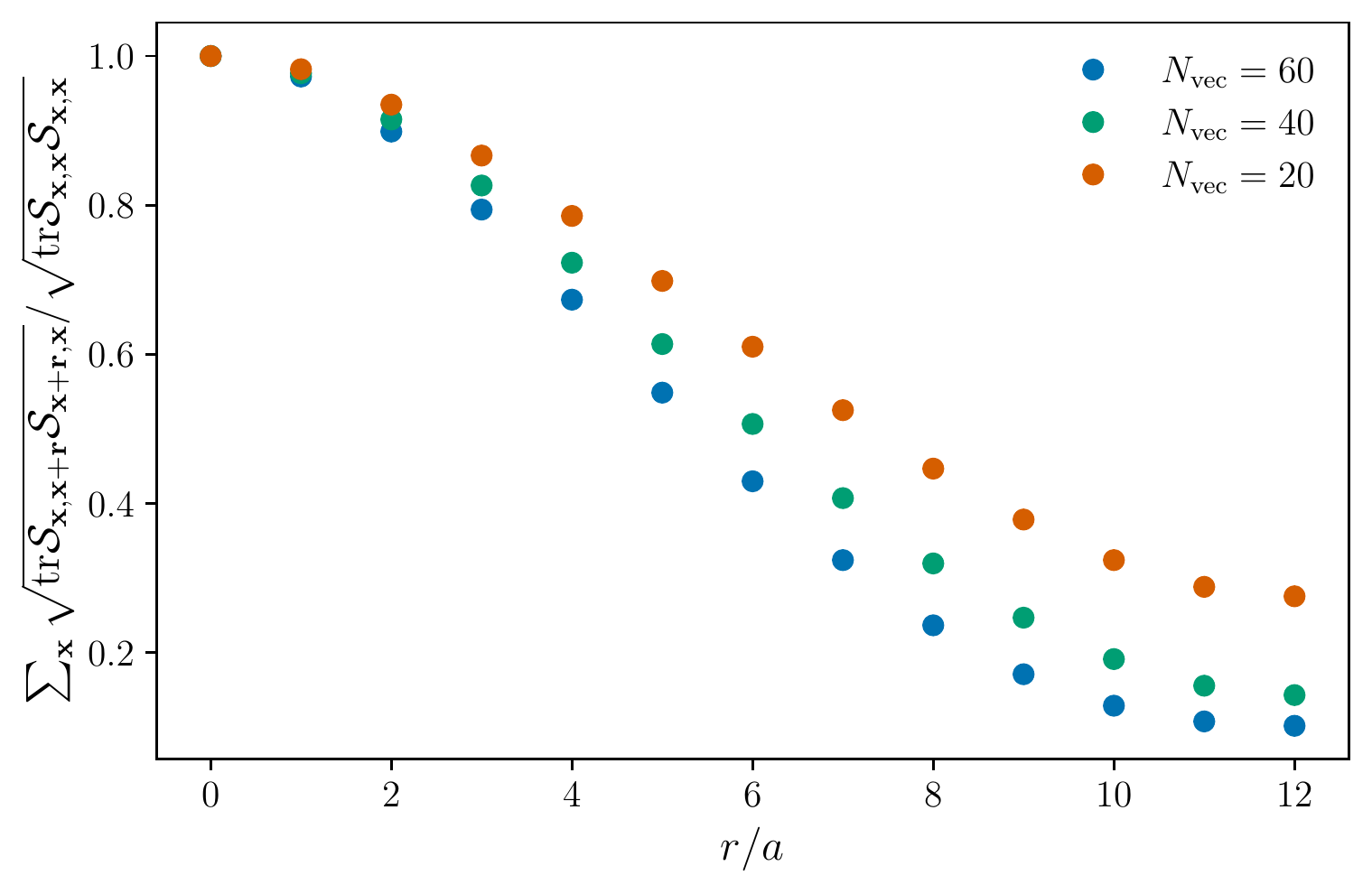}
        \caption{Smearing profile of the distillation operator as a function of $N_\mathrm{vec}$ on ensemble a12m400 with stout-smearing \cite{Morningstar:2003gk} parameters $\rho=0.1$ and $n=1$.}
        \label{fig:smearing_profile}
    \end{figure}

When choosing the number of eigenvectors $N_\mathrm{vec}$ of the gauge-covariant Laplacian for the smearing matrix, \cref{eq:smearing_matrix}, one has to find a compromise between the statistical error and anticipated smearing radius on the one side and the computational cost and memory requirement on the other. One method to get an idea of the required number of eigenvectors is to look at the spatial distribution of the distillation operator as suggested in ref.~\cite{HadronSpectrum:2009krc}. This spatial distribution function defined as
\begin{align}
    \Psi(r)=\sum_{\mathbf{x}}\frac{\sqrt{\mathrm{tr}\mathcal{S}_{\mathbf{x},\mathbf{x}+\mathbf{r}}\mathcal{S}_{\mathbf{x}+\mathbf{r},\mathbf{x}}}}{\sqrt{\mathrm{tr}\mathcal{S}_{\mathbf{x},\mathbf{x}}\mathcal{S}_{\mathbf{x},\mathbf{x}}}}\,,
\end{align}
with $r=|\mathbf{r}|$ is shown in \Cref{fig:smearing_profile} for ensemble a12m400 and $N_\mathrm{vec}\in \{20,40,60\}$.
The smearing profile allows one to get an estimate for the smearing radius for a given number of eigenvectors and therefore an idea of the relevant scales in the computation. It is, however, non-trivial how the estimated smearing radius translates into operator overlaps and overall statistical precision of the data. Instead of using the smearing profile as a benchmark for the impact of $N_\mathrm{vec}$ we opt for an empirical approach and study the quality of finite-volume energies extracted via a GEVP as a function of $N_\mathrm{vec}$.

\begin{figure}[hbtp!]
        \centering
        \includegraphics[width=0.95\linewidth]{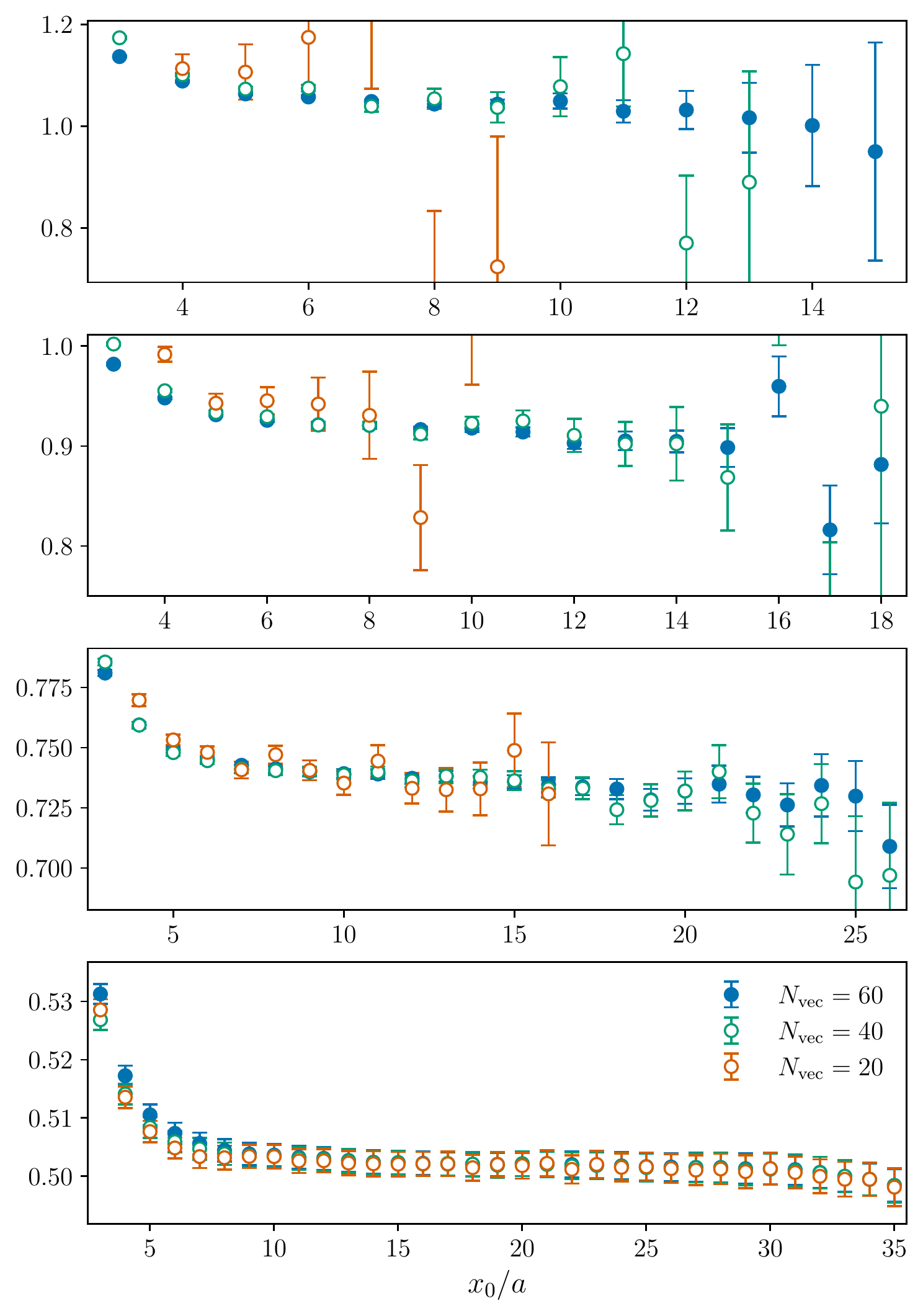}
        \caption{Effective masses from a GEVP with $t_0=2$ for different values of $N_\mathrm{vec}$ for $i=0,1,2,3$ from bottom to top calculated on 77 configurations of ensemble a12m400.}
        \label{fig:effective_masses}
    \end{figure}

In this contribution we restrict our discussion to the repulsive $I=3/2$ $s$-wave channel. In \Cref{fig:effective_masses} we show the effective masses defined by
\begin{align}
    am_\mathrm{eff}(t)=\log \left(\frac{\lambda(t,t_0)}{\lambda(t+1,t_0)} \right)\,,
\end{align}
where $\lambda$ is an eigenvalue obtained by solving a GEVP defined in \cref{eq:GEVP} for given state $i$ and reference time slice $t_0=2$.
From the bottom panel it becomes obvious that the $N_\mathrm{vec}$ has very little impact on the quality of the extracted ground state energy. In line with our expectation a smaller number of eigenvectors corresponds to a larger smearing radius and thus results in slightly better overlap with the ground state which can be seen from the fact that the effective mass settles to a plateau at earlier source sink separations $x_0/a$.
This observation drastically changes for the higher excited states for which a higher number of eigenvectors improves both the statistical quality and the overlap of the GEVP extracted correlator with the desired state. Particularly for the third excited state (top panel in \Cref{fig:effective_masses}) $N_\mathrm{vec}=20$ does not seem to sufice to obtain a reliable estimate of the associated finite-volume energy. We take the fact that the plateaus for all relevant states in this channels agree at moderate source sink separations for both $N_\mathrm{vec}=40$ and $N_\mathrm{vec}=60$ as confirmation that $60$ eigenvectors seems to be a reasonable compromise and proceed with this setup.

\section{Lellouch-Lüscher factors}

\begin{figure}[ht]
        \centering
        \includegraphics[width=0.75\linewidth]{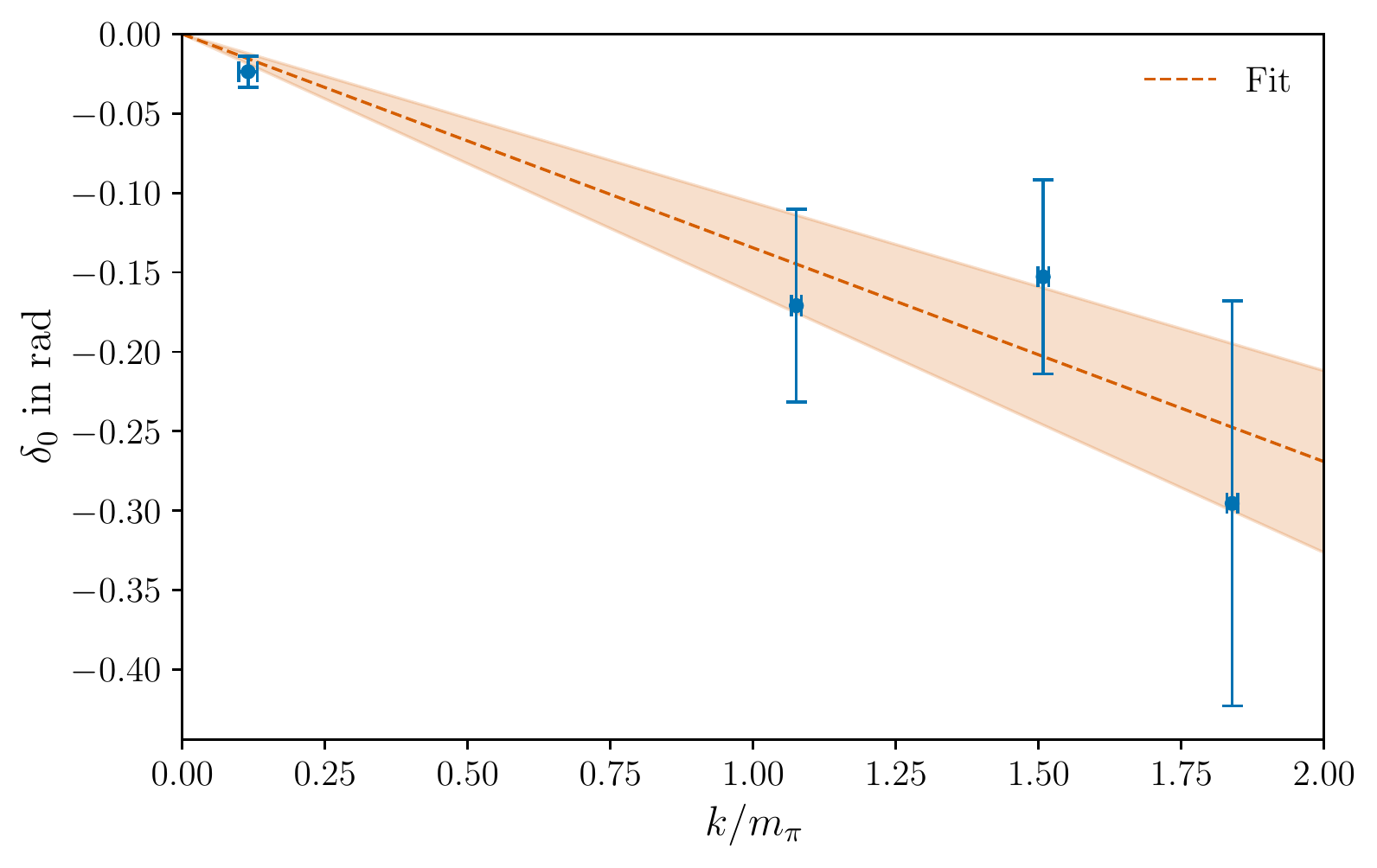}
        \caption{$I=3/2$ $s$-wave scattering phase shifts as a function of $k/m\pi$ together with a linear fit to the data.}
        \label{fig:delta_0}
    \end{figure}

From the finite-volume energies extracted from our preferred data set with $N_\mathrm{vec}=60$ we can derive the corresponding scattering phase shifts via the relation
\begin{align}
    \delta_0(q)=\arctan\left( \frac{\pi^\frac{3}{2}q}{Z_{00}(1;q^2)} \right)\,,\quad q=\frac{k L}{2\pi}\,,
\end{align}
originally derived by Lüscher \cite{Luscher:1990ux,Luscher:1991cf}. The corresponding phase shifts which we obtain for the $I=3/2$ $s$-wave channel are shown in \Cref{fig:delta_0} as a function of $k/m_\pi$, together with a linear fit to the data.
With this model for the scattering phase shift we can get an estimate for the ``Lellouch-Lüscher'' factors relating the finite-volume to the infinite-volume decay amplitudes which are summarized in \Cref{tab:ll-factors}.
    \begin{table}[ht]
        \centering
        \begin{tabular}{S[table-format=1.8]S[table-format=3.6]}
        \toprule
        {$q$} & {F} \\
        \midrule
        0.110(16) & 117(27) \\
        1.0253(87) & 69.84(65) \\
        1.4375(93) & 59.60(41) \\
        1.7530(96) & 80.99(37) \\
        \bottomrule
        \end{tabular}
        \caption{Finite-to-infinite volume proportionality factors $F^2=8\pi \big\lbrace q\frac{\partial \phi}{\partial q}+k\frac{\partial \delta_0}{\partial k}\big\rbrace\frac{E_n^2}{k_{n}^3}$.}
        \label{tab:ll-factors}
    \end{table}

    \begin{figure}[ht]
        \centering
        \includegraphics[width=0.75\linewidth]{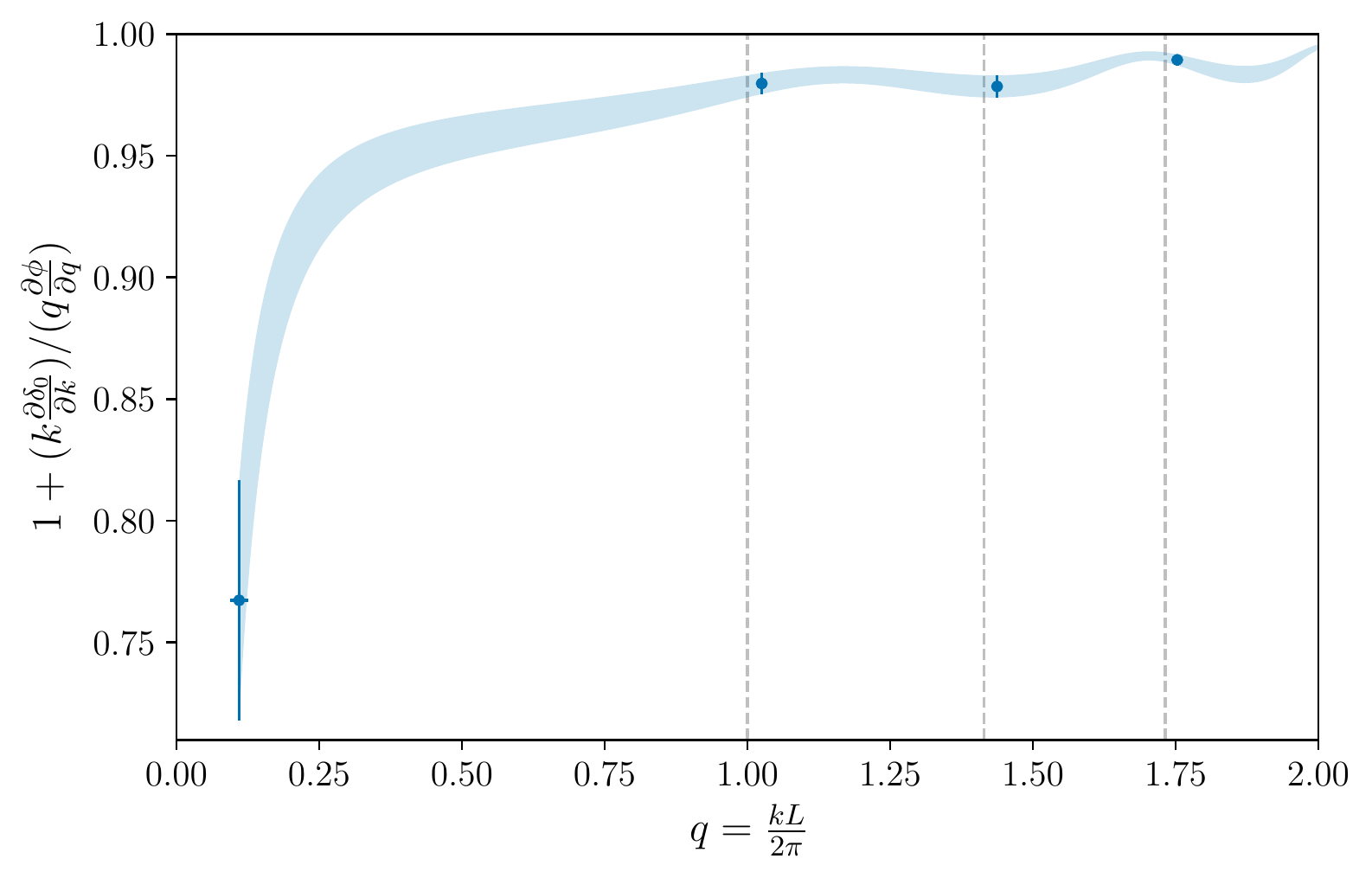}
        \caption{Finite-to-infinite-volume proportionality factors as a function of $q$ divided by their non-interacting counterparts. The shaded blue area is not a fit to the data but an estimate for the expected statistical uncertainty for values of $q$ which are not covered by our data set.}
        \label{fig:rel_LL}
    \end{figure}

To get an idea of the overall impact of our scattering calculation on the proportionality factors we display these factors divided by their non-interacting counterparts in \Cref{fig:rel_LL}. For all four $q$ values in our calculation we see a statistical significant difference from unity highlighting the importance of the scattering analysis for the extraction of hadronic $D$-decay amplitudes.

\section{Conclusions \& Outlook}
In these proceedings we describe our steps towards the first ab-initio calculation of hadronic decays of $D$-mesons. In our simplified setup we focus on the $D\rightarrow K\pi$ decay channel at non-physical quark masses. In order to construct operators which excite $K\pi$ final states with energies close to the $D$-meson mass we make use of the exact distillation method. We use an empirical approach to determine the number of eigenvectors of the gauge covariant Laplacian and found that $N_\mathrm{vec}=60$ is a good compromise for our setup. With the results from a scattering phase shift analysis for the repulsive $I=3/2$ $K\pi$ channel we were able to obtain first results for the ``Lellouch-Lüscher'' proportionality factors which relate the finite-volume matrix elements to the infinite-volume decay amplitudes.

\section*{Acknowledgments}
M.~T.~H. and F.~J.~are supported by UKRI Future Leader Fellowship MR/T019956/1. N.~L. and A.~P. received funding from the European Research Council (ERC) under the European
Union’s Horizon 2020 research and innovation programme under grant agreement No 813942. A.~P., M.~T.~H. and F.~E. are supported in part by UK STFC grant ST/P000630/1. A.~P. and F.~E. also received
funding from the European Research Council (ERC) under the European Union’s Horizon 2020
research and innovation programme under grant agreements No 757646.

This work used the DiRAC Extreme Scaling service
at the University of Edinburgh, operated by the Edinburgh Parallel Computing Centre on behalf of the STFC
DiRAC HPC Facility (www.dirac.ac.uk). This equipment was funded by BEIS capital funding via STFC capital grant ST/R00238X/1 and STFC DiRAC Operations
grant ST/R001006/1. DiRAC is part of the National e-Infrastructure.

The authors acknowledge the open lattice initiative for providing the gauge ensembles. Generating the ensembles the open lattice initiative received support from the computing centres hpc-qcd (CERN), Occigen (CINES), Jean-Zay (IDRIS) and Ir\`ene-Joliot-Curie (TGCC) under projects (2020,2021,2022)-A0080511504 and (2020,2021,2022)-A0080502271 by GENCI as well as project 2021250098 by PRACE and from the DiRAC Extreme Scaling service.

\bibliographystyle{JHEP}
\bibliography{refs}

\end{document}